# Spin gap and superconductivity in the three-dimensional attractive Hubbard model

Raimundo R. dos Santos*

*Departamento de Física, Pontifícia Universidade Católica do Rio de Janeiro, CP 38071, 22452-970 Rio de Janeiro, Brazil*

(May 26, 1994)

We study the phase diagram for the attractive (*i.e.*, negative-$U$) Hubbard model on a simple cubic lattice, through Monte Carlo simulations. We obtain the critical temperature, $T_c$, for superconductivity from a finite-size scaling analysis of the data for the pairing correlations. For fixed on-site attraction, $U$, $T_c$ displays a maximum near the filling factor 0.9, roughly independent of $U$. For fixed filling we estimate the crossover temperature $T^\times(U)$, separating the normal states: metallic and spin-gap. There is also a critical value $U_p$ for pair formation, the magnitude of which seems to be independent of doping. The relevance of these results to the high-$T_c$ oxides is discussed.

One of the most striking normal state properties of the high-$T_c$ cuprate superconductors is the behavior of the uniform magnetic spin susceptibility, $\chi_s$, as the temperature is lowered: Instead of being temperature-independent as in conventional Fermi liquids, $\chi_s$ starts to decrease well above the critical temperature $T_c$, as evidenced by NMR Knight shifts and relaxation rates[1] and by direct susceptibility measurements[2]. This suppression of $\chi_s$ has been associated with the opening of a spin gap at a crossover temperature, $T^\times$, above $T_c$[3-5]. As the temperature is decreased further, and within a certain range of doping, the material becomes superconductor. In the search for a mechanism responsible for superconductivity in these materials, it is therefore instructive to study simplified models displaying the essential features observed, such as the precursor spin-gap phase in the normal state. The attractive Hubbard model (*i.e.*, negative on-site coupling $U$) is believed to display these features[6-8,4]. Early mean-field calculations[6] indicate that local singlet pairs are formed at high temperatures, and that these incoherent pairs condense into a charged superfluid at $T_c$. In two dimensions, Randeria *et al.*[4] have provided numerical evidence to show that the uniform susceptibility of the model is suppressed above the superconducting temperature and that it is proportional to the NMR relaxation rate. Due to their layered structure, one should expect some properties of the cuprates to interpolate between the two- and three-dimensional models. Local fermion pairs may be formed in narrow-band systems due to a local attractive short-ranged effective interaction and have also been invoked to explain a variety of other phenomena; see Ref. 6 for a list of references. More recently, a model for the $CuO_2$ planes with interacting carrier and insulating bands and repulsive interactions has been mapped onto the attractive Hubbard model[9]. In view of all this, a systematic study of the attractive Hubbard model in three dimensions is in order. In particular, there are many aspects such as the behavior of $T_c$ and $T^\times$ with both $U$ and the occupation away from half-filling, that are known at most qualitatively. With this in mind, here we address these questions through Monte Carlo simulations.

The Hubbard Hamiltonian can be written as

$$\mathcal{H} = -t \sum_{\langle i,j \rangle \atop \sigma} \left( c^\dagger_{i\sigma} c_{j\sigma} + \text{H.c.} \right) + U \sum_i \left( n_{i\uparrow} - \frac{1}{2} \right)\left( n_{i\downarrow} - \frac{1}{2} \right)$$
$$-\mu \sum_{i,\sigma} n_{i\sigma} , \qquad (1)$$

where the sums run over sites of a simple-cubic lattice, $\langle i,j \rangle$ denotes nearest neighbor sites, H.c. stands for Hermitian conjugate, and $c^\dagger_{i\sigma}$ ($c_{i\sigma}$) creates (annihilates) a fermion at site $i$ with spin $\sigma$; $U < 0$ is the attractive on-site interaction and $\mu$ is the chemical potential controlling the band-filling. Since the simple-cubic lattice is bipartite, the band is half-filled when the Hamiltonian (1) displays particle-hole symmetry, or $\mu = 0$. In this case, superconducting correlations in the attractive model are equivalent to planar magnetic correlations in the repulsive model[6]. The strong-coupling limit of (1) can be obtained through perturbation theory in the space of doubly occupied states and is equivalent[10,11] to a Heisenberg model in a transverse field proportional to $\mu$.

Here we use a grand-canonical Quantum Monte Carlo simulation; see Refs. 12–15 for details. The imaginary time is discretized through the introduction of $M$ "time" slices separated by an interval $\Delta\tau$ such that $\beta \equiv \Delta\tau M$. One should stress that the simulation for the attractive Hubbard model is free from "minus sign" problems[11,14,15]. We calculate quantities such as the equal-time $\mathbf{q} = 0$ local (or $s$-wave) pairing correlation function,

$$P_s(T,L) \equiv \langle \Delta^\dagger \Delta + \Delta \Delta^\dagger \rangle , \qquad (2)$$

where $T$ is the temperature, $L$ is the linear lattice size, and

$$\Delta^\dagger = \frac{1}{\sqrt{N_s}} \sum_i c^\dagger_{i\uparrow} c^\dagger_{i\downarrow} , \qquad (3)$$

and the uniform magnetic susceptibility

$$\chi_s = \frac{1}{N_s} \sum_{ij} \int_0^\beta d\tau \, \langle m_i(\tau) m_j(0) \rangle \,, \qquad (4)$$

with

$$m_i(\tau) = e^{\tau \mathcal{H}} [n_{i\uparrow} - n_{i\downarrow}] e^{-\tau \mathcal{H}} \,. \qquad (5)$$

The dependence of the pairing correlation function with the system size can be extracted through finite-size scaling (FSS) arguments[16]. For an infinite three-dimensional system one expects a superconducting transition within the $XY$-model universality class, with pairing correlations decaying algebraically at the critical temperature $T_c$. For a finite system of size $L$, one can assume the following FSS ansatz for its associated uniform Fourier transform[11]:

$$P_s(T, L) = L^{2-\eta} F(L/\xi) \,, \qquad (6)$$

where $\xi$ is the correlation length for the infinite system, and $F(z)$ is a scaling function such that $F(z) \to$ const when $L \ll \xi$; in three dimensions[17], $\eta \simeq 0$. At $T_c$, $\xi = \infty$, so that $L^{-2} P_s(T_c, L)$ is a constant independent of lattice size. By plotting $L^{-2} P_s(T, L)$ as a function of $T$ for systems of different sizes, an estimate of $T_c$ can be obtained as the temperature where two successive curves intercept[18].

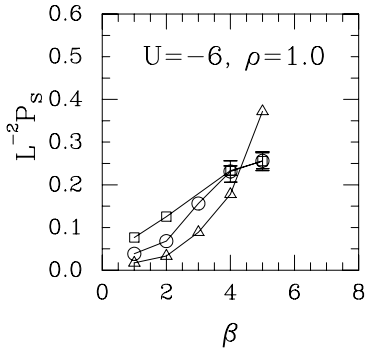

FIG. 1. Size-scaled $\mathbf{q} = \mathbf{0}$ Fourier transform of the $s$-wave pairing correlation function as a function of the inverse temperature, for lattices with $L = 3.17$ (squares), $L = 4$ (circles), and $L = 6$ (triangles), with $U = -6$ and at half-filling. Except where shown, the error bars are smaller than the data points, and the solid lines are guides to the eye.

The clusters used here have $N_s = L_x \times L_y \times L_z$ sites, with periodic boundary conditions; that is, each site is connected with its six nearest neighbors through a hopping term. The simulations were performed on Sun and IBM RISC-6000/525 workstations; a single datum point involves between 500 and 4000 MC sweeps over all time slices and we took $\Delta\tau = 0.125$. In a grand-canonical simulation, for each temperature the chemical potential is adjusted to obtain the desired occupation, $\rho \equiv \langle n \rangle$.

Since we are interested in several values of both $U$ and $\rho$, we had to restrict ourselves to small systems due to our limited computer capabilities. From now on, energies will be expressed in units where the hopping $t = 1$, and we also set the Boltzmann constant $k_B = 1$.

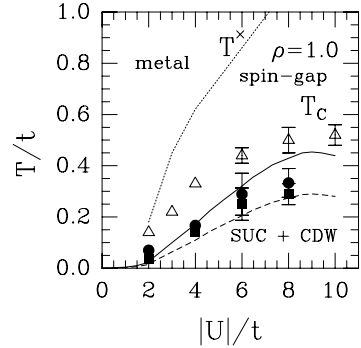

FIG. 2. Critical temperature $T_c$ as a function of the magnitude of the on-site coupling constant for half-filled band; below $T_c$ the system displays both superconductivity (SUC) and charge (CDW) ordering. The results from this work, using $L_3$ (solid circles) and $L_2$ (solid squares) are compared with those obtained from the *repulsive* model: Monte Carlo simulations (open triangles; Ref. 19) and variational calculations (solid and dashed lines; Ref. 20). The dotted line is the crossover temperature $T^\times$ (see text).

We considered lattices with $4 \times 4 \times 2$ and $4 \times 4 \times 4$ sites; but in order to assess possible finite-size effects we have also performed a few runs on a $6 \times 6 \times 6$ lattice. For the $L_x \times L_y \times L_z$ lattices, one may think of several definitions[16] for a mean linear size $L$, such as $L_1 \equiv \sqrt{3/(L_x^{-2} + L_y^{-2} + L_z^{-2})} \simeq 2.83$, $L_2 \equiv 3/(L_x^{-1} + L_y^{-1} + L_z^{-1}) = 3$, and $L_3 \equiv (L_x L_y L_z)^{\frac{1}{3}} \simeq 3.17$. Figure 1 shows the size-scaled pairing correlation, Eq. (2), as a function of the inverse temperature, for $U = -6$ at half-filling and for three different lattice sizes; the data for the $4 \times 4 \times 2$ lattice are plotted assuming $L = L_2 = 3$. One can define the inverse transition temperature, $\beta_c$, as the value where data points for two different-sized systems $(L, L')$ superimpose[18]. This implies $T_c \simeq 0.25$ and $T_c \simeq 0.23$ for (4,3) and (6,4) scalings, respectively; using $L = L_3 = 3.17$ for the smaller lattice, one would get $T_c \simeq 0.3$ from the (4,3) scaling. The definition $L = 3$ for the smaller system then provides estimates for $T_c$ that are closer to the more reliable scaling (6,4) than $L_1$ or $L_3$. Ideally a definite trend would only be detectable through the use of systems larger than $L = 6$, which would become prohibitively expensive in terms of computer time. Taking into account the error bars for the data points, the above criterion implies a typical error $\Delta\beta_c \sim \pm 1$. This procedure is repeated for other values of $U$, to obtain $T_c(U)$ at half-filling. In Fig. 2, the solid symbols repre-sent the critical temperatures obtained from a $(4,\tilde{L})$ scaling, both with $\tilde{L} = L_2$ (squares) and with $\tilde{L} = L_3$ (circles). One should have in mind that *at half-filling* the ordered phase below $T_c(U)$ corresponds to both superconductivity and charge ordering, since the order parameter displays full three-dimensional rotational symmetry[6]. Also, the attractive model at half-filling is equivalent to the *repulsive* model, with the superconducting and charge order parameters becoming the planar (XY) and axial (Z) staggered magnetizations, respectively. In view of this, in Fig. 2 we compare results for $T_c$ from the attractive model with the Néel temperature $T_N$ for the repulsive model obtained from very extensive simulations[19] (within a different extrapolation to estimate $T_N$; open triangles), and from a Gutzwiller-type variational calculation[20] (the solid line is the "bare" result $(T_N(U))$, and the dashed line is an adjustment $((3.83/6.0) \times T_N)$ to fit the mean-field result to that of high-temperature series expansions for the Heisenberg model, according to which[17] $T_N \simeq 3.83 t^2/|U|$). The estimates for $T_c$ using $L_1$ lie below the "normalized" $T_N(U)$ which is probably a lower bound; from now on all quoted estimates for $T_c$ will be based on $L_2$. For weak couplings (*i.e.*, $|U| << t$), the system is in a BCS-like regime; the difference with respect to the standard BCS theory being due to the fact that quasi-particles with any wavevector can pair, not only those close to the Fermi level. Accordingly, the critical temperature is still exponentially small[6], but with a different energy scale: $T_c \sim W \exp(-W/|U|)$, where $W = 3t$ is one half of the bandwidth.

and $U = -1$, the behavior of $\chi_s$ is similar to that of the non-interacting (metallic) case. In contrast, for $U \leq -2$, the uniform susceptibility is suppressed below some temperature $T^\times(U)$. This can be understood in the strong-coupling regime by noticing that local pairs are being formed and that spin excitations necessarily imply pair breaking with an energy cost (gap) of order $|U|$. The formation of local pairs, and the associated spin gap, should be reflected in the magnetic properties: bound singlet pairs must have smaller response to a uniform field than isolated fermions. At intermediate couplings, this behavior can be explained along similar lines, in terms of pairing correlations[4]. Therefore, $T^\times(U)$ represents a crossover temperature separating two normal-state regions: metallic and spin-gap. In Ref. 4, this crossover temperature was defined as the one at which $\chi_s$ deviates from a renormalized Random Phase Approximation. Here we choose a different definition, which follows closely the experimental criterion based on NMR relaxation measurements, namely as the temperature at which $\chi_s$ is maximum; see *e.g.*, the discussion in Ref. 5. The crossover line obtained this way is displayed in Fig. 2. We have compared the data in Fig. 3 with some obtained for the $L = 6$ lattice, and found no significant finite-size effects. Nevertheless, in view of the arbitrariness of this definition, the crossover line obtained is only a crude estimate, and should be taken with care.

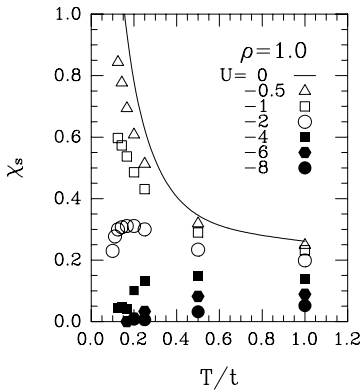

FIG. 3. Uniform susceptibility as a function of temperature at half filling for a simple-cubic lattice with $L = 4$. The symbols refer to the values of $U$ shown, and error bars are smaller than the data points.

In Fig. 3 we present the uniform susceptibility as a function of temperature for the $L = 4$ lattice at half-filling, and for several values of $U$. The solid line is the non-interacting result, $\chi_s^{(0)}$, for the same lattice size; its divergence as $T \to 0$ is due to the finite-size of the system, since $\chi_s^{(0)}$ approaches the Pauli behavior if the $L \to \infty$ limit is taken before $T \to 0$. For $U = -0.5$

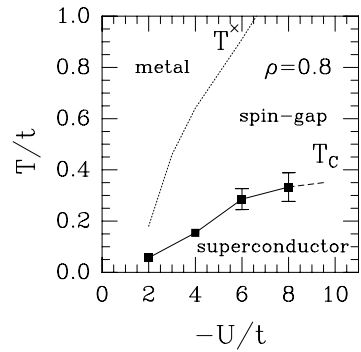

FIG. 4. Critical temperature as a function of the on-site coupling $-U/t$, for $\rho = 0.8$ (solid squares); the solid line is a guide to the eye. The dotted line is the crossover temperature $T^\times$ (see text).

As the occupation is varied, the behavior does not change drastically. For instance, in Fig. 4 we show both $T^\times$ and $T_c$ as functions of $-U$, for an occupation $\rho = 0.8$. Away from half-filling the order parameter is *two*-dimensional, corresponding to superconductivity alone; *i.e.*, charge ordering is lost. While $T^\times$ is roughly the same (within the range of $U$ examined here) as for $\rho = 1$, $T_c$ is slightly higher than for the half-filled case. A plot of $\rho(\mu)$ for the non-interacting $L = 4$ system at zero temperature displays several plateaus; in particular there is one at $\rho = 0.6875$, corresponding to 44 fermions in the system. These plateaux are still present in the interacting case, and are rounded at finite temperatures. This is a finite-size effect that should disappear in the thermodynamic limit, but nonetheless affect the data for $\rho = 0.7$ in these small systems: the uniform susceptibility is strongly suppressed for any $U < 0$. The data for $\rho = 0.6$ are free from these effects.

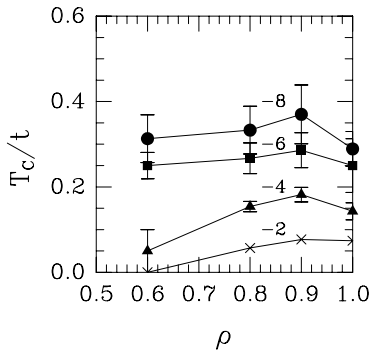

FIG. 5. Critical temperature as a function of the band filling, for the values of $U$ labeling the curves. The solid lines are drawn to guide the eye.

In Fig. 5 we show the dependence of $T_c$ with $\rho$ for several values of $U$. The data are consistent with $T_c$ displaying a maximum around $\rho \simeq 0.9$. For the range of $U$ studied, $T_c$ increases with $|U|$ for fixed occupation, but should eventually decrease in the strong-coupling limit. In the dilute limit (i.e., $\rho \to 0$), $T_c$ should approach zero, for any $U$. But Fig. 5 indicates that the range of fillings for which finite-temperature superconductivity is effective increases with $|U|$.

It is interesting to note[6] that pairs are not formed for any $U$, but below a critical value $U_p$, a precise estimate of which would require further extensive simulations. Nevertheless, by inspection of Fig. 3 one can say that $U_p \in [-2, -1]$, since $\chi_s$ is suppressed for $U = -2$, but not for $U = -1$; this value is quite smaller than the value $U_p \simeq 7.8$ for any $\rho$, predicted within a low-density approximation[6] as the binding energy of the pair. For $\rho = 0.9$, 0.8 and 0.6, $U_p$ lies in the same interval, suggesting that $U_p$ may be insensitive to the occupation. The crossover temperature for $U = -2$ and $\rho = 0.6$ is about 30% smaller than for the other fillings, while for $U \in [-6, -4]$ it seems to be less dependent on the occupation.

In conclusion, we have obtained the phase diagram in the temperature–coupling constant–occupation space for the attractive (i.e., negative-$U$) Hubbard model on a simple cubic lattice. For fixed $U$, the critical temperature for superconductivity displays a maximum at the occupation $\rho \simeq 0.9$. For fixed occupation there are two regimes: weak coupling ($|U| \ll t$), where superconductivity sets in at very low temperatures, from a normal metallic state; and intermediate- to strong-couplings, where superconductivity sets in from a spin-gap phase at higher temperatures. The changeover from a normal metal to a spin-gap phase occurs at a crossover temperature, which does not seem to be very sensitive to the occupation in the range [0.6,1.0], at least for $U \leq -4$.regime. We have also established that the critical value of $|U|$ for pair formation lies in the interval $[-2, -1]$, for all fillings examined. With respect to the cuprates, the existence and origin of the superconducting gap has not been fully settled yet. If the spin gap, which opens above (and not at) $T_c$ in underdoped samples, is the only one present, then describing superconductivity as arising from the condensation of pre-formed pairs as in the model considered here is quite appealing. In this respect, we should comment on a recent suggestion[21] that the spin gap in the attractive model may be irrelevant to the cuprates, as the observed suppression of the uniform susceptibility would be due solely to antiferromagnetic fluctuations. It may be possible to reconcile both views if one considers a Hubbard model with on-site *repulsion* and nearest-neighbor attraction. In this case, the superconducting phase is near a spin-density wave (SDW) instability[6], and SDW fluctuations could influence the magnetic response as suggested. Moreover, the superconducting order parameter in that region has been predicted[6] to have $d$-wave symmetry, in agreement with penetration depth[22], and photoemission[23] studies. Work is in progress to test these ideas.

## ACKNOWLEDGMENTS

The author is grateful to M. A. Continentino for useful discussions. Financial support from the Brazilian Agencies MCT, CNPq, and CAPES is also gratefully acknowledged.